\documentclass[aps,prl,amsmath,amssymb,floatfix,twocolumn,amsmath,superscriptaddress,twocolumn,nofootinbib,tighten,letterpaper]{revtex4-1}

\usepackage{graphicx}
\usepackage{dcolumn}
\usepackage{bm}
\usepackage{longtable}
\usepackage{amsmath}
\usepackage{multirow}

\newcommand{\bvec}[1]{{\mathbf #1}}

\begin{document}

\title{Incompressible Even Denominator Fractional Quantum Hall States in the Zeroth Landau Level of Monolayer Graphene}

\author{Sujit Narayanan}
\affiliation{Department of Physics, Simon Fraser University, 8888 University Drive, Burnaby, British Columbia, V5A 1S6, Canada.}

\author{Bitan Roy}
\affiliation{Max-Planck-Institut f\"{u}r Physik komplexer Systeme, N\"{o}thnitzer Stra. 38, 01187 Dresden, Germany}

\author{Malcolm P. Kennett}
\affiliation{Department of Physics, Simon Fraser University, 8888 University Drive, Burnaby, British Columbia, V5A 1S6, Canada.}

\date{\today}

\begin{abstract}
Incompressible even denominator fractional quantum Hall states at fillings $\nu = \pm \frac{1}{2}$ and $\nu = \pm \frac{1}{4}$ have been 
recently observed in monolayer graphene.  We use a Chern-Simons description of multi-component fractional quantum
 Hall states in graphene to investigate the properties of these states and suggest variational wavefunctions that may describe them.  
We find that the experimentally observed even denominator fractions and standard odd fractions (such as $\nu=1/3, 2/5$, etc.) can be 
accommodated within the same flux attachment scheme and argue that they may arise from sublattice 
or chiral symmetry breaking orders (such as charge-density-wave and antiferromagnetism) of composite Dirac fermions, 
a phenomenon unifying integer and fractional quantum Hall physics for relativistic fermions. We also discuss 
possible experimental probes that can narrow down the candidate broken symmetry phases for the fractional quantum Hall 
states in the zeroth Landau level of monolayer graphene.
\end{abstract}

\maketitle

When graphene is placed in a strong perpendicular magnetic field, a plethora of quantum Hall states are 
observed~\cite{Novoselov2005,Zhang2005,Gusynin2005,Zhang2007,Bolotin2009,Skachko2010,Abanin2013,Young2012,Yu2013,Du2009,Dean2011,Feldman2012,Feldman2013,Amet2015}. 
Interactions among electrons can strongly influence these states when the density is low~\cite{Khveshchenko,HJR,Drut2009,Gorbar2002,Yang2007,Goerbig-review}. 
At the level of the integer quantum Hall effect, the $\nu = 0$ and $\nu = \pm 1$ states are examples of interaction induced Hall 
states~\cite{Skachko2010,Abanin2013,Yu2013,Young2012,Du2009,Dean2011,Feldman2012,Roy2014,Herbut2007a,Herbut2007b,Herbut2008,Semenoff2011,BR-BLG,Barlas2012,Kharitonov2012}, 
consistent with the scenario of sublattice or chiral symmetry breaking (CSB) orders~\cite{Herbut2007a,Herbut2007b,Herbut2008,Kharitonov2012,Roy2014}. 
Investigations of the fractional quantum Hall (FQH) effect in graphene~\cite{Bolotin2009,Du2009,Skachko2010,Dean2011,Feldman2012,Feldman2013,Amet2015} have 
revealed unusual patterns of fractions~\cite{Feldman2012} and unexpected behaviour in a tilted magnetic field \cite{Amet2015,Balram2015b}.

Particularly notable is the very recent observation of incompressible even-denominator fractional quantum Hall (EDFQH) states at $\nu = \pm \frac{1}{2}$ and $\nu = \pm \frac{1}{4}$~\cite{Zibrov2018}. These EDFQH states had not been previously observed in monolayer graphene, although EDFQH states have been seen previously in higher Landau levels (LLs) in single-component systems at $\nu = \frac{5}{2}$ in GaAs~\cite{Willett1987}, and at $\nu = \frac{3}{2}$ and $\frac{7}{2}$ in ZnO~\cite{Falson2015}. In multi-component systems, there have been observations of EDFQH states in bilayer graphene at fractions corresponding to $n = 1$ orbital wavefunctions~\cite{Ki2014,Zibrov2017,Li2017} and at fractions of $\nu = \frac{1}{2}$~\cite{Suen1992,Eisenstein1992,Liu2014a,Liu2014b} and $\frac{1}{4}$~\cite{Luhrman2008,Shabani2009},
corresponding to $n=0$ orbital wavefunctions in systems with multiple layers or sub-bands.

One of the distinguishing feature of the FQH effect in monolayer graphene that there are four isospin components in the zeroth LL corresponding to two valley and two spin degrees of freedom~\cite{Apalkov2006,Khveshchenko2007,Goerbig2007,TokeJain2007a,Modak2011,Peterson2014,Frassdorf2018,deGail2008,Beugeling2010,Shibata2009,Sodemann2014,Balram2015a}. In addition, due to strong electronic interactions in graphene (such as onsite Hubbard repulsion), these states cannot be assumed to be spin polarized. This allows for more degrees of freedom than in systems that have previously demonstrated EDFQH states at $\nu = \pm \frac{1}{2}$ and $\nu = \pm \frac{1}{4}$, and a wide variety of possible states need to be considered in composite fermion or Chern-Simons theories. Previous theoretical studies of the integer 
Quantum Hall states at $\nu = 0$ and $\nu = \pm 1$ that take into account filled LLs \cite{Herbut2007a,Herbut2007b,Herbut2008,Kharitonov2012,Roy2014,Feshami2016,Lukose2016,DeTar2016,DeTar2017} have 
inferred a preference for CSB orders due to strong LL mixing.  Calculations based on this idea have shown good agreement with experiment~\cite{Herbut2008,Roy2014}. 
In Ref.~\cite{Roy2014} two of us argued for the presence of a canted antiferromagnet (CAF) for $\nu = 0$ and charge-density-wave (CDW) order 
with a small component of Neel antiferromagnetism (AFM) at $\nu = 1$. Hence we suggest that CSB may also occur for FQH states with $0 < |\nu| < 1$.

In this paper we make use of the framework for the Chern-Simons theory of multi-component FQH states in graphene 
in the presence of symmetry breaking orders~\cite{Khveshchenko2007,Modak2011,Cai2013} to investigate possible 
composite fermion wavefunctions for the observed EDFQH states. In the $n = 0$ LL of graphene, sublattice and 
valley degrees of freedom are equivalent in the absence of sublattice-symmetry breaking orders. 
We start from a chirally symmetric background and allow for the possibility of dynamical symmetry breaking in the FQH states.

The possibility of incompressible EDFQH states in monolayer graphene was noted in Ref.~\cite{Modak2011}. However, there are numerous ways 
to realize such fractions. Our approach to identifying candidate variational states is as follows. First, we consider flux attachment 
schemes that give either $\nu = \frac{1}{2}$ or $\nu = \frac{1}{4}$.  Second, we note that Zibrov {\it et al.}~\cite{Zibrov2018} observed 
that the magnetic fields at which EDFQH states are observed, some odd denominator fractions coexist with them, while other 
fractions disappear or weaken (with some sample dependence). For example, at fields where $\nu = \frac{1}{2}$ was observed, 
both $\nu = \frac{1}{3}$ and $\nu = \frac{2}{5}$ were also present, and in their sample B, $\nu = \frac{3}{7}$ 
and $\nu = \frac{4}{9}$ were also unaffected. For $\nu = \frac{1}{4}$, they found that $\nu = \frac{1}{5}$ and $\nu = \frac{2}{7}$ 
were mostly present at the same field, whereas $\nu = \frac{2}{9}$ and $\nu = \frac{3}{11}$ were generally not, but were seen at higher 
and lower magnetic fields. We use these observations to winnow out candidate states by postulating that for filling fractions close to 
the EDFQH states with the same flux attachment scheme are the most likely states to be seen at the same magnetic field.  We also determine the other fractions that naturally arise 
from the flux attachments that give rise to EDFQH states and compare with the experimental observations to narrow down the possible states that might give rise to EDFQH effects.  

Our main result is that we identify candidate variational wavefunctions for the observed EDFQH states, which are summarized in Tables~\ref{tab:MMS1nuhalf} and \ref{tab:MMS2nuquarter}. We observe that the majority of these candidate states show CSB in the form of either a CDW or AFM. In light of this result and the role that chiral symmetry plays in the integer quantum Hall effect in the zeroth LL~\cite{Roy2014}, we suggest that CSB is likely a unifying phenomenon for both regular and composite Dirac fermions in the zeroth LL in monolayer graphene. We discuss experiments that can be used to test this idea and to discriminate between potential orderings for a given flux attachment scheme.

The effective low energy Hamiltonian for graphene is $H = H_+ \oplus H_-,$ which acts on eight component spinors $\Psi = \left[\Psi_{\bvec{K}}, \Psi_{-\bvec{K}}\right]^T$, 
where for $\tau = \pm$, $\Psi^T_{\tau\bvec{K}} = [ u_{\uparrow},v_{\uparrow},u_{\downarrow}, v_{\downarrow}](\tau \bvec{K})$, $\pm \bvec{K}$ 
label the two valleys and $u_\sigma$($v_\sigma$) is the amplitude on the $A$($B$) sublattice of graphene's
honeycomb lattice with spin projection $\sigma = \uparrow,\downarrow$. In the absence of symmetry breaking orders $H_{\pm}$ can be written as (setting $\hbar$, $v_F$ = 1)
\begin{eqnarray}
	H_{\pm} & = & \pm I_2 \otimes \sigma_1 \; D_1	- I_2 \otimes \sigma_2 \; D_2,
\end{eqnarray}
where $D_i = -i\partial_i - eA_i$, $\bvec{A}$ is the vector potential. We label the valley-spin configurations $(\bvec{K}\uparrow), (\bvec{K}\downarrow), (-\bvec{K}\uparrow), (-\bvec{K}\downarrow)$ by $\alpha = 1, 2, 3, 4$, respectively. We can thus write the kinetic part of the Hamiltonian as~\cite{Modak2011}
\begin{equation}
H=  \sum_\alpha \Psi_\alpha^\dagger \left(\pm \sigma_1 D_1 - \sigma_2 D_2\right) \Psi_\alpha, \nonumber 
\end{equation}
where $\Psi_\alpha^\dagger  = \left(u_\alpha^\dagger, v_\alpha^\dagger\right)$. We introduce the transformation $ \Psi_\alpha = e^{i\Phi_\alpha} \tilde{\psi}_\alpha$, where $\tilde{\psi}_\alpha$ is a composite fermion field and
\begin{equation} 
\Phi_\alpha = K_{\alpha\beta} \int d\bvec{r}^\prime {\rm arg}(\bvec{r} - \bvec{r}^\prime) \rho_\beta(\bvec{r}^\prime). \nonumber
\end{equation}			
Under this transformation
\begin{equation} 
\Psi_\alpha^\dagger \left(\pm \sigma_1 D_1 - \sigma_2 D_2\right) \Psi_\alpha  \longrightarrow
   \tilde{\psi}_\alpha^\dagger \left( \pm \sigma_1 \tilde{D}_1 - \sigma_2 \tilde{D}_2\right) \tilde{\psi}_\alpha, \nonumber
\end{equation}
where $\tilde{D}_{1,2} = D_{1,2} - a^\alpha_{1,2}$, with Chern-Simons field
\begin{equation}
 \bvec{a}^\alpha = K_{\alpha\beta} \int d\bvec{r}^\prime g(\bvec{r} - \bvec{r}^\prime) 
 \rho_\beta(\bvec{r}^\prime); \quad\, g(\bvec{r}) = \frac{\hat{\bvec{z}} \times \bvec{r}}{r^2}. \nonumber 
\end{equation}	
 Requiring the $\tilde{\psi}_\alpha$ to be fermionic constrains the values of $K$ so that $K_{\alpha\beta}$ is integer-valued with  $K_{\alpha \beta} = K_{\beta\alpha}$ and $K_{\alpha\alpha}$ even~\cite{Rajaraman1997}. In the composite fermion picture the filling fraction $\nu_\alpha$ for species $\alpha$ of composite fermion is related to the densities $\rho_\alpha = \psi^\dagger_\alpha \psi_\alpha = \Psi^\dagger_\alpha \Psi_\alpha$ by~\cite{Modak2011}
\begin{equation}~\label{eq:rhoK}
	\frac{\rho_\alpha}{\nu_\alpha} = \frac{\rho}{\nu} - K_{\alpha\beta} \: \rho_\beta.
\end{equation}

\begin{table}[t!]
	\begin{tabular}{|c|c|c|c|}
		\hline
		$(k,m,n)$ & $(\nu_1, \nu_2, \nu_3, \nu_4)$  & $(C, F, N)$ & Other fractions  \\
		\hline
		(1,2,1) & (1, 0, 1, 0) & (0, 1, 0) & \multirow{2}{*}{ $\frac{\bvec{1}}{\bvec{3}}$, $\frac{\bvec{2}}{\bvec{5}}$, 
                                                       $\frac{\bvec{3}}{\bvec{7}}$, $\frac{\bvec{4}}{\bvec{9}}$} \\
		        & (1, 0, 0, 1) & (0, 0, 1) &  \\
		        & (0, 1, 1, 0) & (0, 0, -1) & \multirow{2}{*}{$\frac{7}{13}$, $\frac{\bvec{5}}{\bvec{9}}$, $\frac{\bvec{4}}{\bvec{7}}$, $\frac{5}{11}$} \\
	                & (0, 1, 0, 1) & (0, -1, 0) &  \\
                        \hline
		(1,1,2) & (1, 1, 0, 0) & (1, 0, 0) & \multirow{2}{*}{$\frac{\bvec{1}}{\bvec{3}}$, $\frac{\bvec{2}}{\bvec{5}}$, 
                                                       $\frac{\bvec{3}}{\bvec{7}}$, $\frac{\bvec{4}}{\bvec{9}}$} \\
	                & (0, 0, 1, 1) & (-1, 0, 0) & \\
		        & (1, 1, 1, 0) & (1, 0, 0) & \multirow{2}{*}{$\frac{7}{13}$, $\frac{\bvec{5}}{\bvec{9}}$, $\frac{\bvec{4}}{\bvec{7}}$, $\frac{5}{11}$} \\
		        & (1, 1, 0, 1) & (1, 0, 0) & \\
		        & (1, 0, 1, 1) & (-1, 0, 0) &  \\
		        & (0, 1, 1, 1) & (-1, 0, 0) & \\
                        & (1, 1, 2, 0) & (1, 0, 0) & \\
                        & (1, 1, 0, 2) & (1, 0, 0) & \\
                        & (1, 1, 1, 2) & (1, 0, 0) &  \\
                        & (1, 1, 2, 1) & (1, 0, 0) & \\
                        & (1, 2, 1, 1) & (-1, 0, 0) & \\
                        & (2, 1, 1, 1) & (-1, 0, 0) & \\
                        & (1, 1, 0, 3) & (1, 0, 0) & \\
                        & (1, 1, 3, 0) & (1, 0, 0) & \\
                        & (0, 3, 1, 1) & (-1, 0, 0) & \\
                        & (3, 0, 1, 1) & (-1, 0, 0) & \\
		\hline
	\end{tabular}
\caption{Parameters for possible $\nu = \frac{1}{2}$ states. Other fractions that can occur for the same $(k, m, n)$ are indicated.  Fractions observed in Ref.~\cite{Zibrov2018} are indicated in {\bf bold}.  Note that when the order parameters take values $\pm 1$ these correspond to the same phase since $C$, $F$ and $N$ represent 
	Ising-like orders.
}~\label{tab:MMS1nuhalf}
\end{table}

We also have the following relations between the composite fermion densities and order parameters
\begin{eqnarray}~\label{eq:orderparameters}
1 &=& {\displaystyle \frac{\rho_{1} + \rho_{2} + \rho_{3} + \rho_{4}}{\rho}}, \:\:
C = {\displaystyle \frac{\rho_{1} + \rho_{2} - \rho_{3} - \rho_{4}}{\rho}},  \nonumber \\
F &=& {\displaystyle \frac{\rho_{1} - \rho_{2} + \rho_{3} - \rho_{4}}{\rho}}, \:\: 
N = {\displaystyle \frac{\rho_{1} - \rho_{2} - \rho_{3} + \rho_{4} }{\rho}}, 
\end{eqnarray}
where $C$ represents CDW, $F$ ferromagnetism and $N$ easy axis Neel order. We parametrize the $K$ matrix as
\begin{eqnarray}~\label{eq:Kmatrix}
K = \left(\begin{array}{cccc} 
      2k_{1} & m_{1} &  n_{1} & n_{2} \\ 
      m_1 & 2k_{2} & n_3 & n_4 \\
      n_1 & n_3 & 2k_3 & m_2 \\
		  n_2 & n_4 & m_2 & 2k_4 
\end{array} \right), 
\end{eqnarray}
and consider the following simplification that $k_1 = k_2$, $k_3 = k_4$, $n = n_1 = n_2 = n_3 = n_4$, so that flux attachment is the same within a valley (sublattice), but not necessarily the same as the other valley (sublattice). We combine Eqs.~(\ref{eq:rhoK}) and (\ref{eq:orderparameters}) to get the set of equations
\begin{equation}
M \left(\begin{array}{c} 1 \\ C \\ F  \\ N \end{array}\right)=\frac{1}{\nu}\left(\begin{array}{c} \nu_{*} \\ \nu_C  \\ \nu_F \\ \nu_N \end{array}\right),
	\label{eq:matrixeqn}
\end{equation}
where we introduce the following quantities
\begin{eqnarray}
	\nu_\ast & =   \nu_{1} + \nu_{2} + \nu_{3} + \nu_{4}, \:\:\:\:
	\nu_C & =   \nu_{1} + \nu_{2} - \nu_{3} - \nu_{4}, \nonumber \\
  \nu_F & =   \nu_{1} - \nu_{2} + \nu_{3} -\nu_{4},  \:\:\:\: 
  \nu_N & =   \nu_{1} - \nu_{2} - \nu_{3} + \nu_{4}, \nonumber 
\end{eqnarray}
and $M$ is written out in full in the Supplementary materials~\cite{Supplementary}. The entries of the matrix $K_{\alpha\beta}$ specify the flux attachment scheme. In the framework of Modak {\it et al.} this corresponds to a variational wavefunction of the form (omitting Gaussian 
factors)~\cite{Modak2011}
\begin{eqnarray}~\label{eq:variationalwf}
	\Psi\left(\{z^\alpha\}\right) & = & {\mathcal P}_{\rm ZLL}\left[\prod_{\alpha = 1}^4 \Phi_{\nu_\alpha}\left(u_1^\alpha, \ldots, 
	u_{N_\alpha}^\alpha\right)\right] \\
	& \times& \prod_{i < j}^{N_\alpha} \left(z_i^\alpha - z_j^\alpha\right)^{2k_\alpha}   \:\:
	\prod_{i,j,\alpha,\beta; \alpha \neq \beta}^{N_\alpha,N_\beta} \left(z_i^\alpha - z_j^\beta\right)^{K_{\alpha\beta}}, \nonumber 
\end{eqnarray}
where for the $N_\alpha$ particles of species $\alpha$, $z_i^\alpha = x_i^\alpha - iy_i^\alpha$ are the complex coordinates 
for the $i^{\rm th}$ particle, $\Phi_{\nu_\alpha}$ is the wavefunction for $\nu_\alpha$ filled LLs of species $\alpha$ and 
${\mathcal P}_{\rm ZLL}$ indicates projection into the zeroth LL (ZLL).  Different parameterizations of 
the $K$ matrix correspond to different variational wavefunctions. We consider parameterizations of increasing complexity and search for solutions of Eq.~(\ref{eq:matrixeqn}) which have either $\nu = \frac{1}{2}$ or $\nu = \frac{1}{4}$.

We use the information about which fractions are seen at the same magnetic field as the EDFQH states to constrain flux attachment schemes that may give rise to these states~\cite{Zibrov2018}. In particular, we postulate that states with the same parameterization of the $K$ matrix are more likely to be robust at the same field, since they differ only in the occupation of composite fermion LLs but not in the nature of the flux attachment. We also expect that states which can be specified with the fewest number of independent entries in the $K$ matrix are the most likely to occur and focus on these as candidate variational states.

We first consider the Toke-Jain states~\cite{TokeJain2007a}. The simplest construction of the $K$ matrix is when all elements are equal, i.e.
$2k = 2k_1 = 2k_3 = m = m_1 = m_2 = n$ and parametrized by a single parameter, $k$. This leads to the Toke-Jain sequence of states:	$\nu = \nu_*/(2k\nu_* + 1)$~\cite{TokeJain2007a}, yielding the sequence $\frac{1}{3}$, $\frac{2}{5}$, $\frac{3}{7}$, $\frac{4}{9}, \ldots$ for $k=1$.
They are always odd denominator states (except in the limit $\nu_* \to \infty$, for which $\nu \to 1/(2k)$ and we expect a compressible composite fermion state~\cite{Modak2011,HLR}) and hence are not candidates for EDFQH states.

We next consider more general states with $k = k_1 = k_3$ and $m = m_1 = m_2$, which are labeled by the \emph{triplet} $(k,m,n)$. 
A simple limit is when $m = 2k$ but $n \neq 2k$, so the flux attachment is of the form $(k,2k,n)$ and specified by two independent 
parameters, $k$ and $n$. One can show  that the allowed fractions for such states are \cite{Modak2011}
\begin{equation}
 \nu = \frac{\nu_\ast + \left(k - n/2\right)\left(\nu_*^2 - \nu_C^2\right)}{1 + 2 \; k \; \nu_* + 
\left(k^2 - n^2/4 \right)\left(\nu_*^2 - \nu_C^2\right)},
\label{eq:MMSIanu}
\end{equation}
and the order parameters can be expressed in simple analytic forms~\cite{Supplementary}. A second class of two parameter states can be obtained by assuming $n = 2k$ and $m\neq 2k$ in which case the flux attachment is of the form $(k,m,2k)$~\cite{Supplementary}. We find that the EDFQH state at $\nu = \frac{1}{2}$ can be described in terms of these two types of flux attachments, but they are insufficient to describe the EDFQH state at $\nu = \frac{1}{4}$.

There are many different triplets $(k,m,n)$ which can give rise to EDFQH states at $\nu = \frac{1}{2}$.  However, if we apply the 
condition that these triplets should also give rise to the fractions $\nu = \frac{1}{3}, \frac{2}{5}, \frac{3}{7}, \frac{4}{9},$ then 
we find that this restricts us to $(k,2k,n)$ states with $k=1$ and $n=1$, i.e. (1,2,1)  states and $(k,m,2k)$ states with $k=1$ and $m=1$, 
i.e. (1,1,2) states. For the (1,2,1) combination we also expect to see incompressible states 
at  $\nu = \frac{7}{13}, \frac{5}{9}, \frac{4}{7}$, and $\frac{5}{11}$ and similarly for (1,1,2). 
The $\nu = \frac{5}{9}$ and $\frac{4}{7}$ states were observed by Zibrov {\it et al.}~\cite{Zibrov2018} but were 
weaker at the fields where the $\nu = \frac{1}{2}$ state was observed. The twenty states that meet these criteria are 
listed in Table~\ref{tab:MMS1nuhalf}. All the (1,1,2) states all have $C \neq 0$, $N = 0$, $F = 0$ while all the (1,2,1) states all have $C = 0$ and 
either $N$ or $F$ non-zero.

\begin{table}[t!]
\begin{tabular}{|c|c|c|c|}
\hline
$(k,m,n)$ & $(\nu_1, \nu_2, \nu_3, \nu_4)$  & $(C, F, N)$ & Other fractions\\
\hline
(2, 3, 2) & (1, 1, 0, 0) & (1, 0, 0) & 
\multirow{2}{*}{
$\frac{\bvec{1}}{\bvec{5}}, \frac{\bvec{2}}{\bvec{9}}, 
\frac{\bvec{3}}{\bvec{13}}, \frac{\bvec{2}}{\bvec{7}},
\frac{\bvec{4}}{\bvec{9}}$} \\
          & (0, 0, 1, 1) & (-1, 0, 0) & \\ \hline
(2, 2, 3) & (1, 0, 1, 0) & (0, 1, 0) & 
\multirow{4}{*}{
$\frac{\bvec{1}}{\bvec{5}}, \frac{\bvec{2}}{\bvec{9}}, 
\frac{\bvec{3}}{\bvec{13}}, \frac{\bvec{2}}{\bvec{7}},
\frac{\bvec{4}}{\bvec{9}}$} \\
          & (1, 0, 0, 1) & (0, 0, 1) & \\
          & (0, 1, 1, 0) & (0, 0, -1) & \\
          & (0, 1, 0, 1) & (0, -1, 0) & \\
\hline
\end{tabular}
\caption{Parameters for candidate $\nu = \frac{1}{4}$ states. Other fractions that can occur for the same $(k, m, n)$ are indicated.  Fractions observed in Ref.~\cite{Zibrov2018} are indicated in {\bf bold}.
}~\label{tab:MMS2nuquarter}	
\end{table}

For the incompressible state at $\nu = \frac{1}{4}$ we first considered states with flux attachment in the form $(k,2k,n)$ and $(k,m,2k)$ and found possibilities with $(k,m,n) = $ 
$(2,4,3)$, $(2,3,4)$ or $(3,6,1)$ as listed in the supplementary materials \cite{Supplementary}. For $k=2$ it is easy to find $(k,2k,n)$ states at the fractions $\nu = \frac{1}{5}, \frac{2}{9}, \frac{3}{13}, \frac{4}{9}$, which are seen in Ref.~\cite{Zibrov2018}, while for $k=3$ and $n=1$ one finds the fractions $\nu = \frac{1}{7}$ and $\frac{2}{13}$ which are not seen in Ref.~\cite{Zibrov2018}, instead of $\nu = \frac{1}{5}$ and $\frac{2}{9}$. However, neither of the combinations $(k,2k,n)$ or $(k,m,2k)$ above
support states at the experimentally observed fraction $\nu = \frac{2}{7}$.

Hence, we consider more general  states with $m \neq 2k$ and $n \neq 2k$, which depend on the three parameters $(k,m,n)$. We solved the 
equations for the filling fractions and order parameters~\cite{Supplementary}, but were not able to find compact analytic forms for their solutions. Noting that the $k=2$ states appear to be more promising for $\nu = \frac{1}{4}$ than the $k=3$ states, we found the following 
combinations in addition to $(2,4,3)$ and $(2,3,4)$ that can give rise to a $\nu = \frac{1}{4}$ EDFQH state: $(2,0,3)$, $(2,1,3)$, $(2,2,3)$, $(2,3,0)$, $(2,3,1)$, $(2,3,2)$, and $(2,3,3)$. When we investigate the above combinations of $(k,m,n)$ to see which combinations also allow for FQHE states at $\nu = \frac{1}{5}$ and $\nu = \frac{2}{7}$, three prominent candidates emerge: $(2,2,3)$, $(2,3,2)$ and $(2,3,3)$.  All three combinations can also have $\nu = \frac{2}{9}$ states, but only the $(2,3,3)$ combination also allows for a $\nu = \frac{3}{11}$ state. Given that the $\nu = \frac{3}{11}$ state disappears at fields at which the $\nu = \frac{1}{4}$ state is observed, we eliminate the $(2,3,3)$ combination, leaving $(2,2,3)$ and $(2,3,2)$ as competing flux attachment schemes. The parameters for these candidate $\nu = \frac{1}{4}$ states are listed in Table~\ref{tab:MMS2nuquarter}.

The $(2,3,2)$ combination has $C\neq 0$, with $F = 0$, $N = 0$, while the $(2,2,3)$ combination has $C = 0$ and allows for either $F \neq 0$ or $N \neq 0$. 
We observed that the $\nu = \frac{2}{7}$ state is quite robust when the $\nu = \frac{1}{4}$ state forms and is similar to the $\nu = \frac{1}{4}$ state in that only one of 
$C$, $N$, or $F$ is non-zero when it occurs.  In contrast, the $\nu = \frac{1}{5}$ and $\nu = \frac{2}{9}$ states have $|C| = |N| = |F| = 1$, and appear to be weaker at the fields where the $\nu = \frac{1}{4}$ EDFQH is observed.

Based on the idea that fractions that coexist with EDFQH states at the same magnetic field are likely to have the same $K$ matrix, but different fillings of Dirac composite fermion LLs, we suggest that the likely candidate variational wavefunctions for $\nu = \frac{1}{2}$ have $(k,m,n) = (1,1,2)$ or $(1,2,1)$ and those
for $\nu = \frac{1}{4}$ have $(k,m,n)$ as $(2,2,3)$ or $(2,3,2)$. Even within 
this limited set of flux attachments there is still three-fold degeneracy associated with the pattern of symmetry breaking orders present in the states, 
as shown in Tables~\ref{tab:MMS1nuhalf} and \ref{tab:MMS2nuquarter}. In order to discriminate further, we need information about the nature of the 
broken symmetries in the various EDFQH states.  Note that $C$ and $N$ are CSB orders and therefore cause strong LL mixing.  
As a result, the onset of CDW and AFM orders may cause the system to lower its energy by pushing all 
filled LLs of composite Dirac fermions further down in energy. Hence we expect any FQH state with 
$C \neq 0$ or $N\neq 0$ to be energetically superior to
those with $F \neq 0$. Such states can be expected to arise in graphene due to electron-electron interactions. 
The pattern of symmetry breaking realized in any particular sample will depend on the relative strength of various finite range components of the Coulomb interaction.

Zibrov {\it et al}. \cite{Zibrov2018} noted that there was a sublattice gap in their experiments, the size of which was correlated with the 
magnetic field at which EDFQH states were seen. They proposed that the EDFQH states are associated with a phase transition from a partially sublattice polarized (PSP) to a
CAF phase. Within the variational states we consider this would correspond to a transition from a state with $C \neq 0$ to one with spin ordering. 
A more general variational state than we have considered here might be achieved by taking linear combinations of states of the form of 
Eq.~(\ref{eq:variationalwf}) with the same $(k,m,n)$ but different $(\nu_1,\nu_2,\nu_3,\nu_4)$. These might give ways to realize PSP or CAF states. 
On the other hand, experiments by Amet {\it et al.}~\cite{Amet2015} reported that the FQHE states in the $n=0$ LL do not show appreciable 
change in a tilted magnetic field, leading them to conclude that the state is possibly spin polarized, which would favour $F \neq 0$.  
However, as noted in Ref.~\cite{Roy2014}, the  order parameters in $\nu = 0$ states (believed to be a CAF) can be relatively 
insensitive to even quite strong parallel fields. Thus it may be possible to have both $F \simeq 0$ and relatively little sensitivity to tilted fields.

We suggest that measurement is the best way to resolve the ambiguity of the nature of the broken symmetry in the EDFQH states. In 
the case of CDW order, sublattice resolved STM measurements could determine the presence of non-zero $C$ at EDFQH states, 
and the spin ordering (either $F$ or $N$) could potentially be probed with spin resolved STM.  Such information could pare 
down the possible states quite significantly. Additionally, studies of edge states via tunnelling measurements could provide 
additional constraints on possible orders~\cite{Zibrov2018}. Investigation of the excitation spectra for 
different possible states might also provide ways to discriminate between different orders. 
The recent construction of a multicomponent Abelian Chern-Simons theory in a functional integral approach is a promising step in this direction \cite{Frassdorf2018}.

The multicomponent states we consider here are considerably more complex in their flux attachment than the standard 
sequence of FQHE states that have been proposed for monolayer graphene but actually show many of the same fractions (e.g. $\frac{1}{3}$, $\frac{2}{5}$, $\frac{3}{7}$, 
$\frac{4}{9}$, \ldots). This observation raises questions about the nature of states that have been observed in graphene 
previously~\cite{Feldman2012,Feldman2013,Amet2015} and whether these do indeed belong to families with the simplest flux attachment. 
Finally we note that the flux attachments for the states at $\nu = \frac{1}{4}$ are related to those found for $\nu = \frac{1}{2}$ by $j \to j+1$ for $j=k,m,n$
and that the symmetry breaking orders flip, i.e. $C \leftrightarrow F, N$.  
This switching of CSB (CDW or AFM) orders is reminiscent of the 
transition from CAF to CDW in going from $\nu = 0$ to $\nu = 1$ \cite{Roy2014} 
and suggests that there is a hierarchical splitting of degenerate composite Dirac fermion LLs occuring within the zeroth LL.

In summary, we propose candidate wavefunctions for the recently observed incompressible EDFQH states at $\nu = \frac{1}{2}$ and $\nu = \frac{1}{4}$. The possibilities we uncover indicate that the zeroth LL in graphene may harbor even more richness in possible electronic states than 
previously anticipated. We urge additional experimental efforts to uncover the nature of these unusual states which may help to pin down the patterns of broken symmetry 
FQH states in graphene.

M. P. K. and S. N. were supported by NSERC and M. P. K. acknowledges the hospitality of the Max-Planck Institute for the Physics of Complex Systems in Dresden while a portion of this work was completed.

\end{document}